\documentclass[a4paper]{article}

\RequirePackage{latexsym}
\RequirePackage[colorlinks,citecolor=blue,urlcolor=blue,linkcolor=blue]{hyperref}
\usepackage{amsmath}

\begin{document}
\begin{center}
 {\Large\bf Unitarity in quantum cosmology: symmetries protected and violated} \\
\vskip 0.3 cm
{\large Sachin Pandey\footnote{Email: sp13ip016@iiserkol.ac.in} and Narayan Banerjee\footnote{Email: narayan@iiserkol.ac.in}} \\
\vskip 0.3 cm
Department of Physical Sciences, Indian Institute of Science Education and Research Kolkata, Mohanpur - 741 246, WB, India

\end{center}

\vskip 0.3 cm

\begin{center}
{\bf  Abstract}
\end{center}

This work deals with the violation or retention of symmetries associated with the self-adjoint extension of the Hamiltonian for homogeneous but anisotropic Bianchi I cosmological model. This extension is required to make sure the quantum evolution is unitary. It is found that the scale invariance is lost, but the Noether symmetries are preserved.


\vskip 0.3 cm


\vskip 0.8 cm 

\section{Introduction}
Wheeler-DeWitt scheme of quantization\cite{wheeler, dewitt} of cosmological models was believed to have been plagued with non-unitarity in anisotropic cosmological 
models\cite{pinto-neto, alvarenga}. Very recently, quite a few examples were shown where anisotropic cosmological models, quantized in the Wheeler-DeWitt scheme, do have 
unitary evolution\cite{sridip, sridip2, sridip3, sridip4, sachin, fabris, barun}. This unitarity is achieved by a suitable operator ordering. In fact a theorem has recently 
been proved\cite{sridip5} which shows that at least for homogeneous models, it is always possible to  have a self-adjoint extension for the Hamiltonian and thus a unitary 
evolution for the system. \\

The purpose of this paper is to look at the price for the self-adjoint extension, in terms of symmetry. We shall look at two aspects, one is the Noether symmetry and the other
being the scale invariance. \\

Noether symmetry has been enormously useful in the context of gravity. In classical gravity, Noether symmetry can lead to finding exact solutions for the system\cite{abhik}, 
helps in understanding the nature of dark energy\cite{basil, jamil} and certainly finds conserved quantities in the system. \\

Noether symmetry finds a very important role in quantum cosmological models. Capozziello and Lambiase\cite{capo-grg} showed that if a Noether symmetry exists in a minisuperspace 
cosmology, it actually acts as a selection rule to determine whether the Hartle condition\cite{hartle} is satisfied and the corresponding classical behaviour can be recovered. In 
the context of extensions of general relativity, either with a non-minimal coupling of a scalar field to geometry or by including higher order derivatives in the relevant 
action, Capozziello, DeLaurentis and Odintsov\cite{capo-epjc} discussed the role of Noether symmetry quite extensively. The crucial result obtained in this work is that the existence of Noether 
symmetry leads to a classification of singularities, which are essentially the points where these symmetries are broken. For a comprehensive review on the status of the use of 
Noether symmetries, both in classical and quantum aspects of gravity, we refer to the work of Capozziello and DeLaurentis\cite{capo-ijgmmp}. \\

Scale invariance on the other hand, if it exists, is considered to be a very useful property of a transformation of variables. However, very recently it has been proved by Pal 
and Grinstein\cite{sridip-scale} that hermiticity and scale invariance are clearly incompatible. \\

As the self-adjoint extension in quantum cosmological models involve transformation of variables, it is thus a requirement to check the fate of these symmetries, and forms the 
motivation of the present work. \\ 

We shall deal with one example, the Bianchi I cosmological model which is the simplest anisotropic cosmological model, but this brings out the associated physical content quite 
comprehensively. \\

In section 2, we briefly describe the self-adjoint extension which makes the evolution unitary in the case of a Bianchi I model. In the next two sections, we discuss about 
the Noether symmetry and the scale invariance respectively. In the last section, we make some concluding remarks. 

\section{Bianchi I spacetime}

This example has been already worked out in an earlier work\cite{sridip}. We narrate the important steps here so as to make the presentation self sufficient.

We start with the action

\begin{equation}\label{action}
A = \int_M d^4x\sqrt{-g}R+2\int_{\partial M} d^4x\sqrt{h}h_{ab}K^{ab}+\int_M d^4x\sqrt{-g}P ,
\end{equation}
in a four dimensional space-time manifold M. R is the Ricci Scalar, $K^{ab}$ is extrinsic curvature and $h^{ab}$ is induced metric on the boundary $\partial M$. The first two 
terms correspond to the gravity sector and the third term is due to a  perfect fluid which is taken as the matter constituent of the universe, $P$ being the pressure of the fluid. 
We have chosen our units such that $16\pi G =1$. The second term in the action will not contribute in the Euler-Lagrange equations as there is no variation in the boundary.\\

Bianchi I metric is given as
\begin{equation}\label{metric}
ds^2=n^2(t)dt^2-a^2(t)dx^2-b^2(t)dy^2-c^2(t)dz^2.
\end{equation}

With this metric, the gravity sector of the action can be written as 
\begin{equation}
\label{action-grav1}
\mathcal{A}_g=\int dt \bigg[-\frac{2}{n}[\dot{a}\dot{b}c+\dot{b}\dot{c}a+\dot{c}\dot{a}b]\bigg].
\end{equation}

Here an overhead dot implies a differentiation with respect to the cosmic time $t$. A transformation of variables as

\begin{eqnarray}
\label{trans-1}
a(t)=e^{\beta_0+\beta_++\sqrt{3}\beta_-}, \\
b(t)=e^{\beta_0+\beta_+-\sqrt{3}\beta_-}, \\
c(t)=e^{\beta_0-2\beta_+},
\end{eqnarray}

will make the Lagrangian in equation (\ref{action-grav1}) look like,

\begin{equation}
\label{lagrangian}
\mathcal{L}_g = -6\frac{e^{3\beta_0}}{n}[\dot{\beta}_0^2-\dot{\beta}_+^2-\dot{\beta}_-^2].
\end{equation}

The canonically conjugate momenta are defined as usual as $p_i = \frac{\partial {\mathcal{L}_g}}{\partial x_i}$. The Hamiltonian for the gravity sector looks like

\begin{equation}\label{hamgrav}
 H_{g}=-n\exp(-3\beta_{0})\left\{\frac{1}{24}\left(p_{0}^{2}-p_{+}^{2}-p_{-}^{2}\right)\right\}.
\end{equation}

For the fluid sector, the practice is to adopt the Schutz formalism of expressing the fluid properties like density and pressure in terms of some thermodynamic variables and 
then effect a set of canonical transformations. The method is described comprehensively in ref. \cite{sridip, barun}. The relevant action is

\begin{equation}
\label{action-fluid}
\begin{split}
\mathcal{A}_{f}&=\int dt \mathcal{L}_{f}\\&= \int dt \left[n^{-\frac{1}{\alpha}}e^{3\beta_{0}}\frac{\alpha}{\left(1+\alpha\right)^{1+\frac{1}{\alpha}}}\left(\dot{\epsilon}+
\theta\dot{S}\right)^{1+\frac{1}{\alpha}}e^{-\frac{S}{\alpha}}\right],
\end{split}
\end{equation}
 where $\epsilon, \theta$ and $S$ are the thermodynamic quantities (Schutz variables), and $\alpha$ is a constant that connects the density ($\rho$) and pressure ($P$) as 
 $P = \alpha \rho$. \\

As the metric components do not depend on spatial coordinates, the spatial volume integrates out as a constant and will not participate in the subsequent calculations. Also, 
the boundary term is ignored as that does not contribute to the variation of the action.\\

The Canonical momenta are defined as $p_{\epsilon}=\frac{\partial\mathcal{L}_{f}}{\partial\dot{\epsilon}}$ and
$p_{S}=\frac{\partial\mathcal{L}_{f}}{\partial\dot{S}}$. Now we effect a set of canonical transformations,
\begin{eqnarray}\label{can-trans}
T&=&-p_{S}\exp(-S)p_{\epsilon}^{-\alpha -1},\\
p_{T}&=&p_{\epsilon}^{\alpha+1}\exp(S),\\
\epsilon^{\prime}&=&\epsilon+\left(\alpha+1\right)\frac{p_{S}}{p_{\epsilon}},\\
p_{\epsilon}^{\prime}&=&p_{\epsilon}.
\end{eqnarray}

The Hamiltonian for the fluid becomes 

\begin{equation}
\label{ham-fl}
H_{f}= ne^{-3\beta_{0}}e^{3\left(1-\alpha\right)\beta_{0}}p_{T}.
\end{equation}

One can now write the net Hamiltonian $H$ as 

\begin{equation}\label{ham-net}
H = H_{g} + H_{f} = n\exp(-3\beta_{0})\left\{-\frac{1}{24}\left(p_{0}^{2}-p_{+}^{2}-p_{-}^{2}\right)+e^{3(1-\alpha)\beta_{0}}p_{T}\right\},
\end{equation}

and proceed to quantize the system by using the Hamiltonian constraint $H=0$ (obtained by varying the action with respect to the lapse function $n$) and raising the variables 
as operators by using the standard commutation relations $[q_{j}, p_{k}] = i\hbar {\delta}_{jk}$. The process yields the Wheeler-DeWitt equation as 

\begin{equation}\label{WDeq}
\left(\frac{\partial^{2}}{\partial \beta^{2}_{0}}-\frac{\partial^{2}}{\partial\beta^{2}_{+}}-\frac{\partial^{2}}{\partial\beta^{2}_{-}}\right)\psi 
= 24\imath e^{3(1-\alpha)\beta_{0}}\frac{\partial \psi}{\partial T}.
\end{equation} 

In writing this equation, a choice of gauge has been made ($n=e^{3\alpha{\beta}_{0}}$). With the usual separability ansatz 
$\psi(\beta_{0},\beta_{+},\beta_{-},T)=\phi(\beta_{0},\beta_{+},\beta_{-})e^{-\imath ET}$ where $E$ is a constant, the equation (\ref{WDeq}) becomes 

\begin{equation}\label{WDeq2}
\left(\frac{\partial^{2}}{\partial \beta^{2}_{0}}-\frac{\partial^{2}}{\partial\beta^{2}_{+}}-\frac{\partial^{2}}{\partial\beta^{2}_{-}}\right)\phi 
=24E\phi e^{3(1-\alpha)\beta_{0}}.
\end{equation}

For the general case $0\leq \alpha \leq 1$, choose another separation of variables as $\phi = \xi (\beta_0) \eta (\beta_{+} , \beta_{-})$ and a coordinate transformation 
as $\chi = e^{-\frac{3}{2}(\alpha-1)\beta_{0}}$. A long but straightforward calculation will yield an equation for $\xi$ as 

\begin{equation}\label{xi}
-\frac{d^{2}\xi}{d\chi^{2}}-\frac{\sigma}{\chi^{2}}\xi = -E^{\prime}\xi,
\end{equation} 

where $E^{\prime}=\frac{32}{3\left(1-\alpha \right)^{2}}E$ and $\sigma$ is composed of the separation constants coming in the process of separation of the functions of 
$\beta_{+}$ and $\beta_{-}$.  This equation indeed has a favourable deficiency index which guarantees the existence of a self-adjoint extension\cite{vaughn} and thus the 
evolution of the system is unitary. \\

The work of Pal and Banerjee\cite{sridip} shows that this transformation even at the classical level gives rise to a Hamiltonian which, when raised to operators, is 
self-adjoint. It deserves mention that often the unitarity is achieved by means of an operator ordering, which is not unique\cite{sridip5}. The most unambiguous 
example would be the case where one can effect a coordinate transformation at the classical level so that the operator ordering is irrelevant at the quantum level. The 
present example is exactly that and this is one good reason for choosing Bianchi I at the outset. For a very brief review of various aspects of factor ordering, we 
refer to \cite{vakili}. 

\section{Preservation of Noether symmetry}

The generator for Lagrangian (\ref{lagrangian}) can be written as: 
 \begin{equation}
 \textbf{X}=b_0\frac{\partial}{\partial\beta_0}+b_+\frac{\partial}{\partial\beta_+}+b_-\frac{\partial}{\partial\beta_-}+
 \dot{b_0}\frac{\partial}{\partial\dot{\beta_0}}+\dot{b_+}\frac{\partial}{\partial\dot{\beta_+}}+\dot{b_-}\frac{\partial}{\partial\dot{\beta_-}},
 \end{equation}

where $b_i (\beta_j)$s ($i, j = 0, +, -$) are to be determined from the Noether symmetry condition
\begin{equation}
\label{noether}
 \pounds_\textbf{X}\mathcal{L}_g=0,
\end{equation}

meaning the Lie derivative of the Lagrangian with respect to $\textbf X$ is zero. \\

This condition yields the set of equations

 \begin{eqnarray}
 \frac{3}{2}b_0+\frac{\partial b_0}{\partial \beta_0}=0,\\
  \frac{3}{2}b_0+\frac{\partial b_+}{\partial \beta_+}=0,\\
   \frac{3}{2}b_0+\frac{\partial b_-}{\partial \beta_-}=0.
 \end{eqnarray}

The solution for this set of equations can be written as

\begin{eqnarray} \label{con1}
 e^{\frac{3}{2}\beta_0}b_0=Q_1=constant, \\
\frac{3}{2}\beta_+b_0+b_+=Q_2=constant, \\
\frac{3}{2}\beta_-b_0+b_-=Q_3=constant.
\end{eqnarray}

A coordinate transformation of the form $\chi=e^{-\frac{3}{2}(\alpha-1)\beta_0}$ transforms the Lagrangian density given in (\ref{lagrangian}) to the form

\begin{equation}\label{ltran}
\mathcal{L}_{g_T}=-\frac{6}{n}[\frac{4\chi^{2\alpha/(1-\alpha)}}{9(1-\alpha)^2}\dot{\chi^2}-\chi^{2/(1-\alpha)}(\dot{\beta}_+^2+\dot{\beta}_-^2)]. 
\end{equation}

It is now required to check whether the Noether symmetry corresponding to $\mathcal{L}_g$ is retained in $\mathcal{L}_{g_T}.$

The corresponding generator for the Lagrangian as in equation (\ref{ltran}) can be written as,
\begin{equation}
\textbf{X}=q_0\frac{\partial}{\partial\chi}+q_+\frac{\partial}{\partial\beta_+}+q_-\frac{\partial}{\partial\beta_-}+\dot{q_0}\frac{\partial}{\partial\dot{\chi}}+
\dot{q_+}\frac{\partial}{\partial\dot{\beta}_+}+\dot{q}_-\frac{\partial}{\partial\dot{\beta}_-},
\end{equation}

where $q_i (\chi,\beta_j)$ ($i,j = 0,+,-$) are to be determined from the Noether symmetry condition 

$\pounds_\textbf{X}\mathcal{L}_{g_T}=0$, which in this case gives following partial differential equations,
 \begin{eqnarray}
 \frac{\alpha}{1-\alpha}\frac{q_0}{\chi}+\frac{\partial q_0}{\partial \chi}=0,\\
 \frac{\alpha}{1-\alpha}\frac{q_0}{\chi}+\frac{\partial q_+}{\partial \beta_+}=0,\\
 \frac{\alpha}{1-\alpha}\frac{q_0}{\chi}+\frac{\partial q_-}{\partial \beta_-}=0.
 \end{eqnarray}

 Solution to above three equations can be given as,
 \begin{eqnarray} \label{con1}
 {q_0}^{1/\alpha}\chi^{1/(1-\alpha)}=Q_A=constant, \\
\frac{q_0 \beta_+}{(1-\alpha)\chi}+q_+=Q_B=constant, \\
\frac{q_0\beta_-}{(1-\alpha)\chi}+q_-=Q_C=constant.
 \end{eqnarray}

It is easy to check that the  solution to Noether symmetry conditions for both Lagrangian match exactly with the identification $b_0
=q_{0}^{1/\alpha}=[\frac{3}{2}(1-\alpha)]^{1/(\alpha-1)}e^{-\frac{3}{2}\beta_0}.$ \\

Thus we conclude that the self-adjoint extension, which requires a transformation of variable at the classical level, preserves the Noether symmetry.

\section{Loss of scale invariance}

We shall now look at the issue of scale invariance. Under a scale transformation,  there is a rule of transformation of the coordinate and time, e.g., if $x=x(t)$ 
is transformed like $\bar{x} = \lambda^{-1/2} x$, $t$ should go like $\bar{t} =\lambda t$, where $\lambda$ is the scale. \\

We now look back at the equation (\ref{xi}), where we easily identify $\chi$ and $T$ as the coordinate and time respectively. So the relevant transformations will be

$\bar{T}=\lambda T$, $\bar{\chi}(\bar{T})=\lambda^{-1/2}\chi(\lambda T)$. Energy  will transform $\bar{E}'=E'/\lambda$ and $\frac{\partial}{\partial{\bar{\chi}}}=
\lambda^{1/2}\frac{\partial}{\partial{\chi}}$. For a comprehensive description of this, we refer to references  \cite{cabo} and \cite{pradhan}.

If we effect this scale transformation, the equation (\ref{xi})takes the form as
 \begin{equation}
 -\lambda\frac{d^2\xi}{d{\chi}^2}-\lambda\frac{\sigma}{{\chi}^2}\xi-=-\frac{E'}{\lambda}\xi.
 \end{equation}

This clearly does not preserve the scale invariance! However, this should not perhaps be considered too costly, as there is indeed an incompatibility between hermiticity 
and scale invariance in general. As already mentioned, this has been proved very elegantly by Pal and Grinstein\cite{sridip-scale}. So this is not at all an artefact of 
anisotropic quantum cosmology and is rather a general feature of self-adjointness.

\section{Conclusion}

It is now known that the alleged non-unitarity of the anisotropic quantum cosmological models is not pathological. Anisotropic models, at least if they are spatially 
homogeneous, are shown to have self-adjoint extension\cite{sridip5}. This rejuvenates the Wheeler-DeWitt scheme of quantization, as now the conservation of probability
is exact and not an approximation. The present work looks at the cost of this extension in terms of symmetry. \\

With the example of the Bianchi I metric, it is quite clearly shown that the self-adjoint extension indeed retains the Noether symmetry. This is important, as finally the models 
have to describe the observed classical universe, and Noether symmetry plays a vital role in picking up the classical behaviour in terms of Hartle condition. For an excellent 
discussion, we refer to the review \cite{capo-ijgmmp}. \\

However, the scale invariance is clearly lost in this game of a self-adjoint extension. So apparently this is the price for securing unitarity and hence the conservation of
probability! As already mentioned, this does not appear to be too much of an unexpected price for securing unitarity, because self-adjointness and scale invariance are not 
compatible to each other\cite{sridip-scale}. \\

We have dealt with a Bianchi-I model. This has been done for the sake of simplicity. But the method, as it stands, is applicable to any model for which the relevant 
transformation of variables is explicitly known. \\

It should also be pointed out that very recently unitarity is restored not only in cosmology but also in other quantum processes in gravity. One crucial example is that in 
Hawking radiation\cite{arpit1, arpit2}. This is a step towards resolving the information loss paradox. \\

Restoring unitarity in quantum gravitational systems will enhance the predictive power and thus have imprints on the possible observational quantities. The present work 
shows that the virtues of the quantum mechanical description is not lost in the process of self-adjoint extension. \\

\vspace{0.4 cm}

{\bf Acknowledgement:} S.P. wants to thank CSIR (India) for financial support. \\

\end{document}